\documentclass[12pt,draftclsnofoot,onecolumn]{IEEEtran}

\usepackage{graphicx}
\usepackage{amsmath}
\usepackage{amssymb}
\usepackage[caption=false]{subfig}
\usepackage[noadjust]{cite}
\usepackage{float}
\usepackage{algorithm}
\usepackage{bibentry}
\usepackage{balance}
\usepackage{algorithm}
\usepackage{algorithmicx}
\usepackage{algpseudocode}
\usepackage{xcolor}
\usepackage{subfig}
\usepackage{wrapfig}
\graphicspath{{img/}}
\usepackage{url}

\begin{document}

\bstctlcite{IEEEexample:BSTcontrol}

\title{UAV Corridor Coverage Analysis with Base Station Antenna Uptilt and Strongest Signal Association
\thanks{This work is supported in part by the NSF award CNS-1939334. }\thanks{S. J. Maeng, \.{I}. G\"{u}ven\c{c} are with the Department of Electrical and Computer Engineering, North Carolina State University, Raleigh, NC 27606 USA (e-mail: smaeng@ncsu.edu; iguvenc@ncsu.edu).}}

\author{\IEEEauthorblockN{Sung Joon Maeng, \.{I}smail G\"{u}ven\c{c}},~\IEEEmembership{Fellow, IEEE}}%

\maketitle

\begin{abstract}
Unmanned aerial vehicle (UAV) corridors are sky lanes where UAVs fly through safely between their origin and destination. To ensure the successful operation of UAV corridors, beyond visual line of sight (BVLOS) wireless connectivity within the corridor is crucial. One promising solution to support this is the use of cellular-connected UAV (C-UAV) networks, which offer long-range and seamless wireless coverage. However, conventional terrestrial base stations (BSs) that typically employ down-tilted sector antennas to serve ground users are not ideally suited to serve the aerial vehicles positioned above the BSs. In our previous work, we focused on studying the optimal uptilt angle of BS antennas to maximize the wireless coverage probability in UAV corridors. However, the association of BSs with UAVs was restricted to the nearest BS association, which limits the potential coverage benefits. In this paper, we address this limitation by considering the strongest BS signal association in UAV corridors, which enables enhanced coverage within the corridor compared to the nearest BS association. The strongest BS association allows UAVs to connect with the second nearest BSs while also accounting for interference from the third nearest BSs. Closed-form expression analysis and simulation results show that the strongest BSs association in UAV corridors yields a superior coverage probability when compared to the nearest BS association.
\end{abstract}


\section{Introduction}\label{sec:intro}
The utilization of drones, also known as UAVs, has gained significant attention due to their various commercial, government, and military applications. UAV base stations (UAV-BSs) are utilized for public safety. During emergencies, when communication infrastructure fails, UAV base stations (UAV-BSs) are employed for public safety communications~\cite{merwaday2015uav}. Additionally, UAVs are instrumental in tasks such as transporting medical supplies and facilitating search and rescue operations~\cite{UNICEF}, and for transporting people using air taxis~\cite{yun2021distributed}. UAVs can be used in carrying boxes for commercial delivery services~\cite{Insider_Intelligence} and capturing and broadcasting live videos~\cite{Forbes}. They also play a vital role in military operations, particularly for surveillance purposes ~\cite{Unmanned_surveillance}. Recognizing the importance of ensuring the safety, security, and operational efficiency of UAVs, the Federal Aviation Administration (FAA) and the National Aeronautics and Space Administration (NASA) have collaborated on the development of a framework known as UAV Traffic Management (UTM)~\cite{FAA2}. Within this framework, the concept of UAV corridors has emerged, aiming to establish designated sky lanes for UAVs to navigate between waypoints, effectively managing the flow of UAV traffic~\cite{bhuyan2022advances,namuduri2022advanced}.

\begin{figure}[t]
    \centering
    \includegraphics[width=0.65\textwidth]{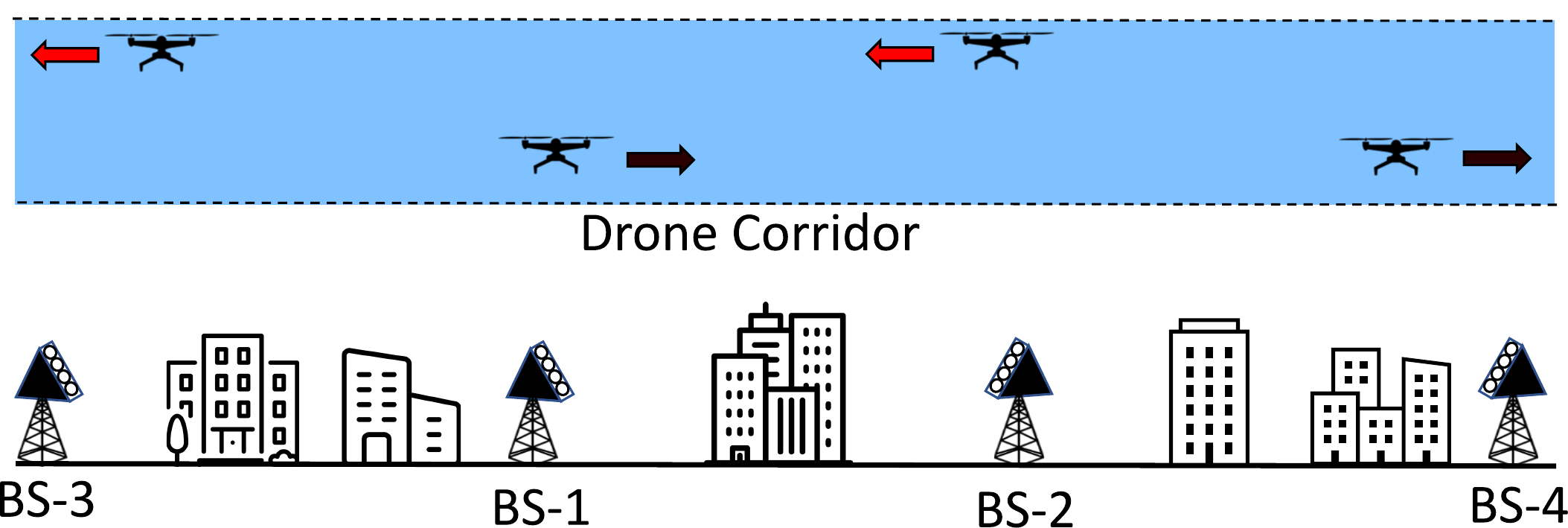}
    \caption{Illustration of the considered UAV corridor scenario. Four BSs are deployed along a straight corridor to serve UAVs.}\label{fig:illu_main}
\end{figure}

In cellular-connected UAVs (C-UAVs), ground BSs can function as wireless service providers for UAVs operating within a corridor. To enable this concept, establishing beyond-visual line of sight (BVLOS) connectivity becomes critical. UAV operators require the ability to remotely monitor and control UAVs even when they are not within direct visual range, and this requires real-time wireless coverage. 
In terrestrial cellular networks, ground BSs use \emph{directional} sectors to serve smartphones and IoT devices located in buildings or outdoor environments, which are situated beneath the BSs. Consequently, the antennas of these BSs are typically down-tilted, a configuration that does not lend itself optimally to serving the UAVs flying in the airspace above the BSs in UAV corridors as illustrated in Fig.~\ref{fig:illu_main}.

The concept of UAV corridors has garnered considerable attention in recent literature. In \cite{muna2021air}, researchers focused on designing air corridors as sky lanes and investigating traffic management and corridor capacity. The optimization of BS locations based on waypoints and antenna array setups to enhance the coverage of UAV corridors has been explored in \cite{singh2021placement}. Stochastic geometry-based analysis~\cite{du2022modeling} and a genetic algorithm-based heuristic method~\cite{chowdhury2021ensuring} were employed to study the improvement in coverage through the use of additional uptilt antennas for C-UAVs. However, these works did not specifically address the unique requirements of UAV corridors. In~\cite{zhang2020radio}, researchers proposed a 3D radio map-based C-UAV path planning approach, aiming to minimize the flying distance with the quality of service (QoS) constraints. The utilization of massive multiple-input-multiple-output (MIMO) in C-UAVs was explored in~\cite{geraci2018understanding,8214963,huang2021massive}. Additionally, the coverage of C-UAV in coordinated multi-point (CoMP) networks taking into account the three-dimensional movement of UAVs has been investigated in \cite{amer2020mobility}, and the coverage of C-UAV networks by considering a three-dimensional antenna pattern has been studied in \cite{lyu2019network}.

This work extends our earlier work in~\cite{maeng2023base}, to consider the strongest BS signal association instead of the nearest BS association. In particular, we analytically derive the outage probability in the corridor as a function of the antenna uptilt angle. We show that the strongest BS association leads to wider coverage in UAV corridors. However, it also introduces a greater complexity compared to the nearest BS association. More specifically, we explicitly account for associations with the second-closest BS and consider interference from the third-closest BS, introduce the concept of the coverage borders based on the SINR threshold to analyze the coverage probability, and identify 6 distinct interference scenarios for individual analysis based on the specific uptilt angles. Note that the unique interference scenarios considered in this manuscript are significantly different and more complicated than those in our prior work~\cite{maeng2023base}, and hence, they require further study for broader applicability considering the closest BS association rule. 
For instance, to analyze the coverage of case~2, additional coverage areas coming from the beam-served area of the BS-2, interference area between BS-2 and BS-3, and coverage borders need to be calculated separately for the closest BS association rule. This requires to define additional angle and height parameters in~(25), which makes the derivation of the closed-form equations more complicated in~(26). On the other hand, in our prior work~\cite[eqn. (16)]{maeng2023base}, only the beam-served area of the BS-1 and the interference area between BS-1 and BS-2 are considered for the analysis, and the corresponding closed-form equation is relatively simple. By employing a combination of closed-form expression analysis and simulation results, we demonstrate that the strongest BS association offers improved coverage compared to the nearest BS association. Additionally, we observe that the outage probability follows a convex curve in relation to the uptilt angles of the antennas.

\section{System Model} \label{sec:system}
We consider the linear UAV corridor networks where ground BSs serve UAVs operating in the air space. These UAVs traverse straight aerial corridors, moving from one waypoint to another. The BSs are strategically deployed along these straight lines, as depicted in Fig.~\ref{fig:illu_main}, between buildings and/or rooftops for advanced air mobility (AAM) scenarios, where air taxis with vertical take-off and landing (VTOL) features may use vertiports located at building rooptops~\cite{yun2021distributed}. To simplify our analysis, we adopt a 2D UAV corridor model, with one axis representing the vertical direction and the other axis representing the horizontal direction relative to the ground on Earth. This model is distinct from terrestrial cellular networks, as UAVs are distributed along a straight aerial corridor. Therefore, the 2D model from the cross-section of the corridor helps us understand the coverage variation of the 3D geometry of the corridor as well.

Unlike visual line of sight (VLOS) flights, which are operated within the pilot’s line of sight, BVLOS flights are flown beyond the visual range. It implies that approaches such as cellular-connected UAVs are necessary to realize BVLOS UAV flights. In other words, due to the height of the antennas at the cellular towers, the communication links between cellular towers and a UAV can still be in LoS, but from the perspective of a UAV pilot, the UAV can be in BVLOS. 
On the other hand, the line-of-sight (LoS) / non-line-sight (NLoS) propagation conditions are decided by the channel link condition between a BS and a UAV depending on the obstacles. Therefore, the BVLOS UAV flights can be possible in either LoS or NLoS propagation conditions.

\begin{figure}[t!]
    \centering
    \subfloat[Case 1 $(h_1 > h_3~\&~h_1 > h_4)$.]{\includegraphics[width=0.4\textwidth]{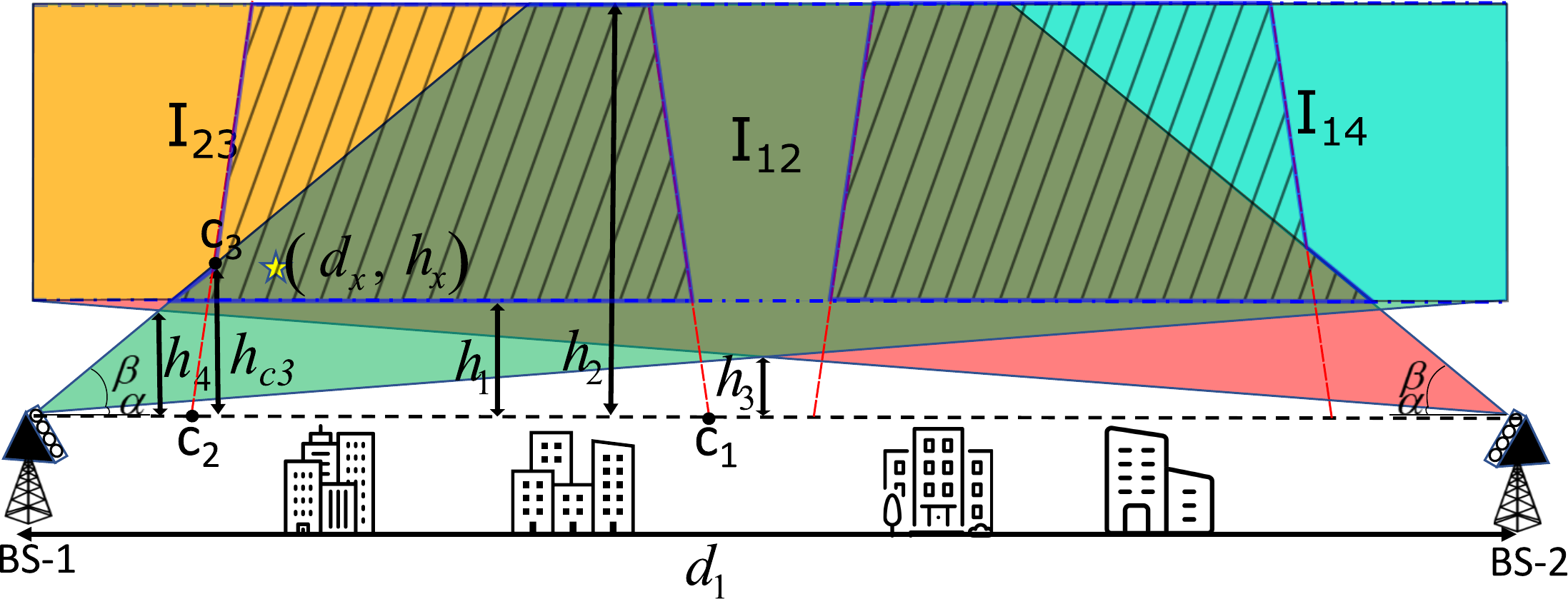}\label{fig:illu_c1}}\vspace{-1.2mm}
    
    \subfloat[Case 2 $(h_1 > h_3~\&~h_1 < h_4)$.]{\includegraphics[width=0.4\textwidth]{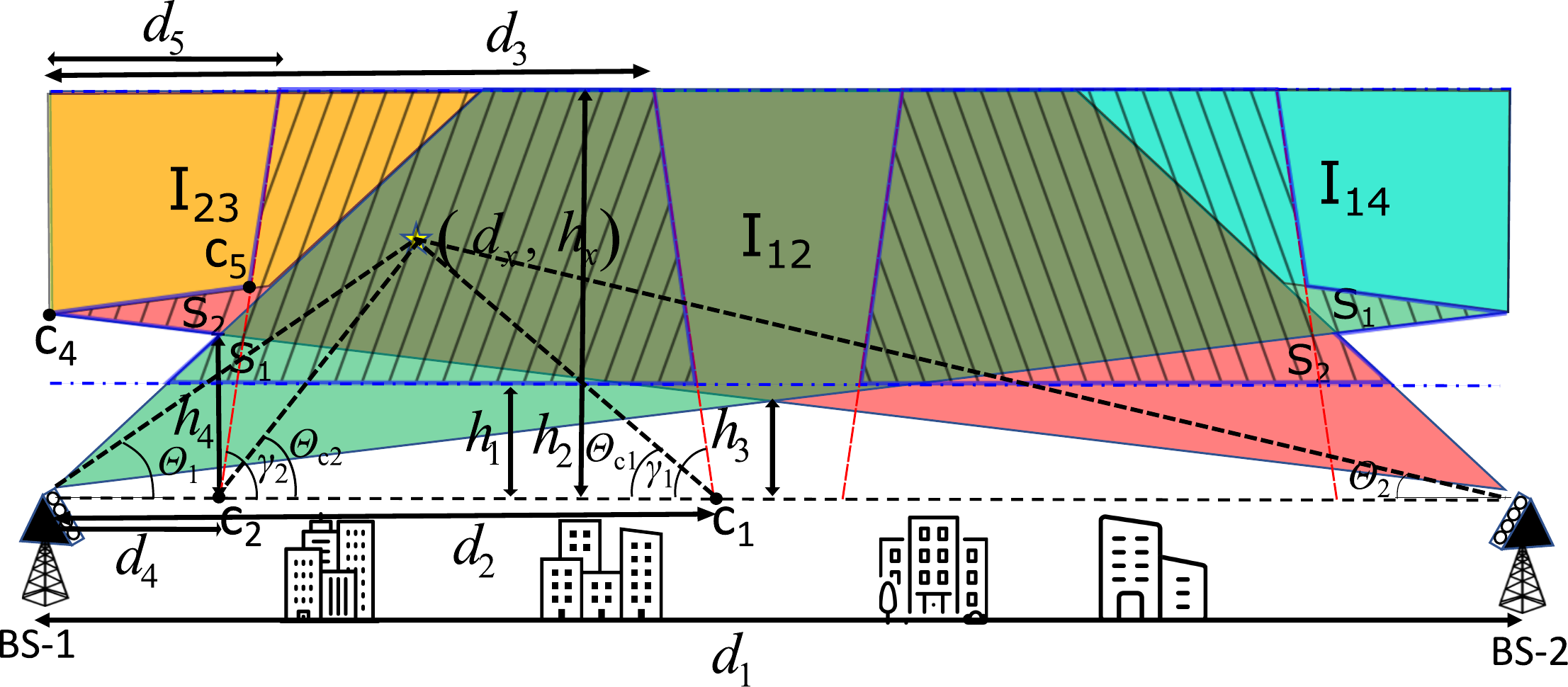}\label{fig:illu_c2}}\vspace{-1.2mm}
    
    \subfloat[Case 3 $(h_1 < h_3~\&~h_1 < h_4~\&~h_2 > h_{\rm c_4})$.]{\includegraphics[width=0.4\textwidth]{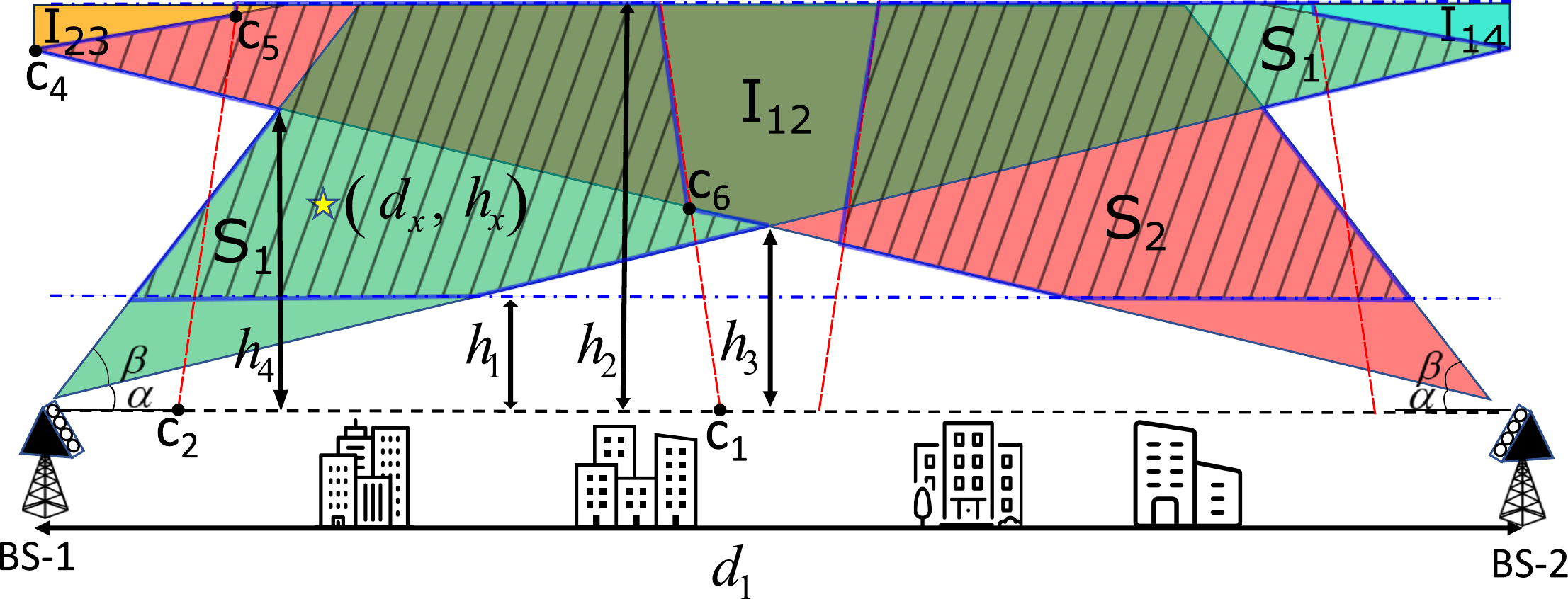}\label{fig:illu_c3}}\vspace{-1.2mm}
    
    \subfloat[Case 4 $(h_1 < h_3~\&~h_1 < h_4 < h_2~\&~h_2 < h_{\rm c_4})$.]{\includegraphics[width=0.4\textwidth]{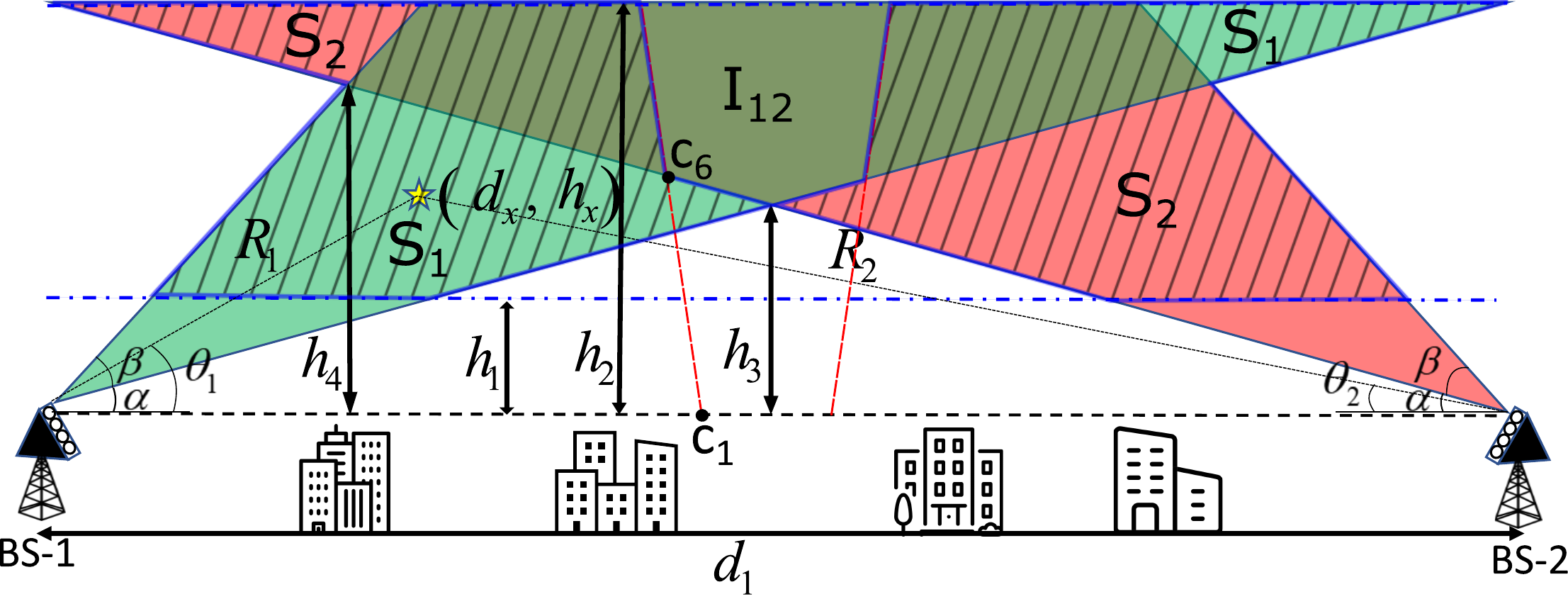}\label{fig:illu_c4}}\vspace{-1.2mm}
    
    \subfloat[Case 5 $(h_1 < h_3 < h_2 ~\&~h_2 < h_4$).]{\includegraphics[width=0.4\textwidth]{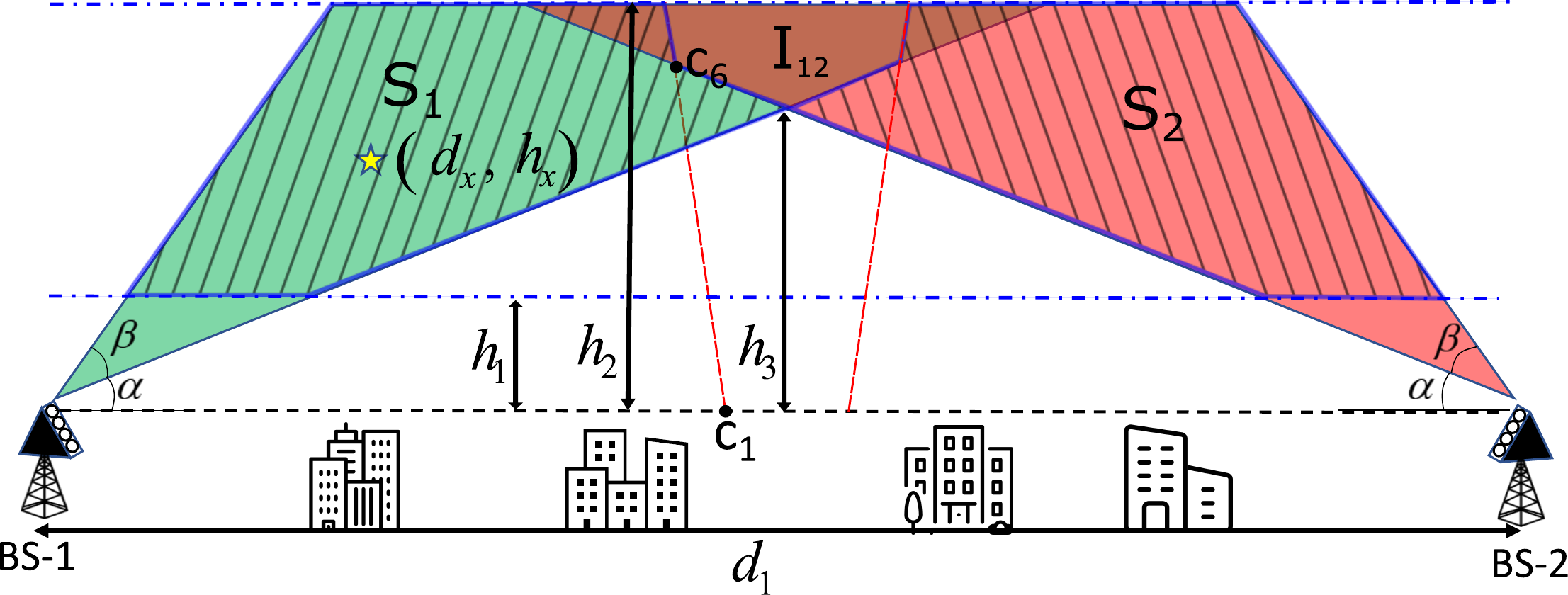}\label{fig:illu_c5}}\vspace{-1.2mm}

    \subfloat[Case 6 $(h_2 < h_3 ~\&~h_2 < h_4$).]{\includegraphics[width=0.4\textwidth]{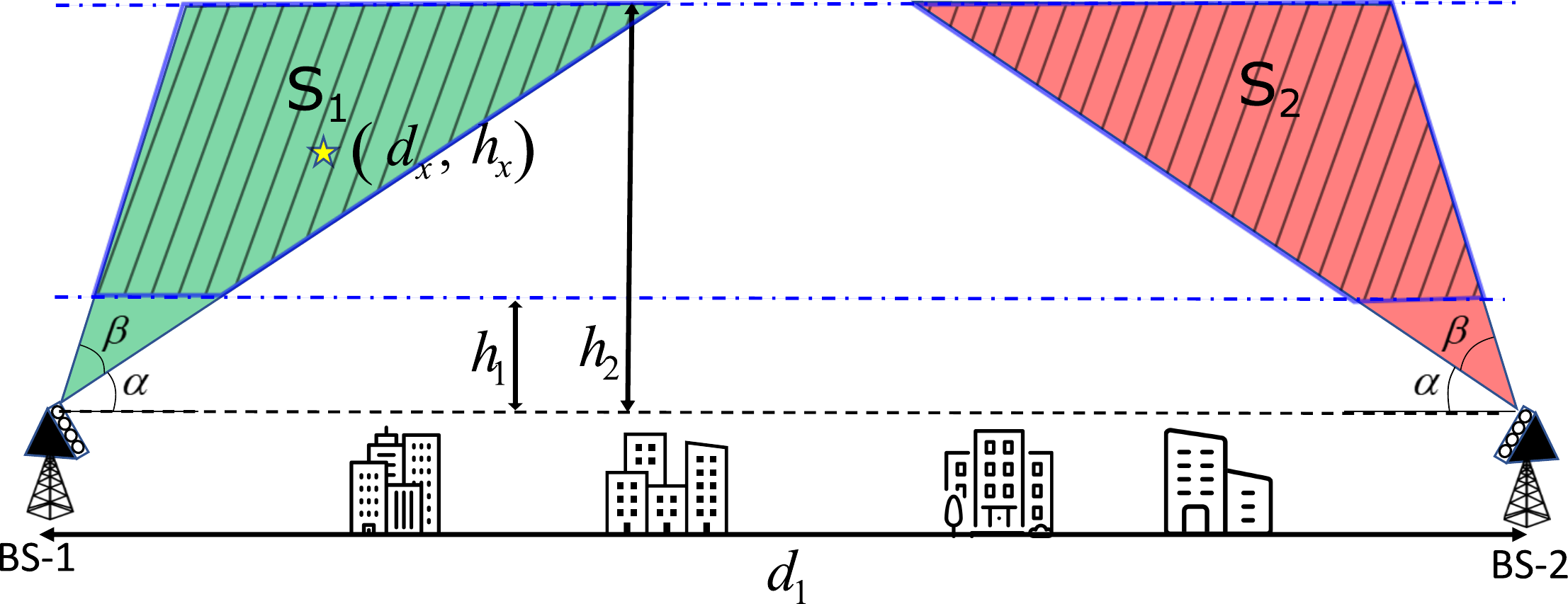}\label{fig:illu_c6}}\vspace{-1.2mm}
    \caption{The illustration of 6 cases as antenna uptilt angle increases. Beam-served areas without interference $\mathsf{S}_1$, $\mathsf{S}_2$, interference area $\mathsf{I}_{12}$, $\mathsf{I}_{23}$, $\mathsf{I}_{14}$ are varied.}
    \label{fig:illu}\vspace{-6mm}
\end{figure}

\subsection{UAV Corridor Modeling}
For analysis simplicity, we assume that the distance between two BSs is the same, and the antenna uptilt angle and the beamwidth of the antenna radiation pattern are identical for different BSs. We consider that UAVs are randomly distributed following uniform distribution in the UAV corridor. The probability density function (PDF) of the horizontal distance ($d_{\rm x}$) and the vertical distance ($h_{\rm x}$) from a BS to a UAV can be expressed as
\begin{align}
    f_{d_{\rm x}}(d_{\rm x})&=\frac{2}{d_1},\;[0<d_{\rm x}<d_1/2],\\
    f_{h_{\rm x}}(h_{\rm x})&=\frac{1}{h_2-h_1},\;[h_1<h_{\rm x}<h_2],
\end{align}
where $d_1$, $h_1$, $h_2$ denote the vertical distance between BSs, the horizontal distance from a BS antenna to the minimum height of the UAV corridor, and the horizontal distance from a BS antenna to the maximum height of the UAV corridor as described in Fig.~\ref{fig:illu}. Since the coverage area is symmetrical due to the identical distance between BSs ($d_1$) and antenna uptilt angle ($\alpha$), and beamwidth ($\beta$), we only consider a half area of the UAV corridor between BS-1 and BS-2 for analysis. The elevation angles from BS-1, BS-2, and BS-3 can be expressed as
\begin{align}\label{eq:theta}
    \theta_1&=\tan^{-1}\left(\frac{h_{\rm x}}{d_{\rm x}}\right),\;\theta_2=\tan^{-1}\left(\frac{h_{\rm x}}{d_1-d_{\rm x}}\right),\nonumber\\
    \theta_3&=\tan^{-1}\left(\frac{h_{\rm x}}{d_1+d_{\rm x}}\right).
\end{align}
Then, the PDF of elevation angle from BS-1 ($\theta_1$) given $h_{\rm x}$ can be derived by the PDF transformation function as~\cite[(4)]{maeng2023base}
\begin{align}
    f_{\theta_1}(\theta_1|h_{\rm x})=\frac{2h_{\rm x}}{d_{1}}\csc^2\theta_1,\;\left[\tan^{-1}\left(\frac{2h_{\rm x}}{d_{1}}\right)<\theta_1<\frac{\pi}{2}\right].
\end{align}

We consider rectangular shape directional beam pattern for analysis, which is written by~\cite{maeng2023base,yi2019unified}
\begin{align}
    g_{\rm i}(\theta_{\rm i})&=
    \begin{cases}\label{eq:bp_model}
    G & \text{if }\alpha<\theta_{\rm i}<\alpha+\beta\nonumber\\
    0 & \text{otherwise}
    \end{cases}~,\\
    \alpha&>0,\; \alpha+\beta<\frac{\pi}{2}~,
\end{align}
where $g_{\rm i}(\theta_{\rm i})$ and $G$ indicate the antenna gain of BS-i and the maximum antenna gain. It is worth highlighting that the main contribution of this paper is the investigation of the strongest base station (BS) association and its impact on coverage improvement, in comparison to the previously studied nearest BS association~\cite{maeng2023base}. By considering the strongest BS association, we introduce a more sophisticated analysis compared to~\cite{maeng2023base} due to the complexity arising from UAVs being served by the BS with the highest signal power among BS-1, BS-2, BS-3, and BS-4. 

Note that beamforming and beam tracking techniques with multiple beams can be considered to mitigate the interference from neighboring beams in the UAV corridor~\cite{maeng2017preamble}. However, this multiple-beam tracking technique requires precise beam tracking to achieve a reliable connection, and it is more complicated for a UAV corridor scenario due to the high mobility of UAVs. For instance, a short-period beam training preamble (pilot) transmission is inevitable to track the high-speed users, which increases transmission overhead. In addition, the number of simultaneous beams is constrained by the number of antennas of base stations (BSs), beamwidth, and orthogonal sequences for the preamble. In this paper, we limit the scope of the study to a single wide sector with antenna uptilt. Note that our analysis can also be applied to beamforming scenarios with ideal beam steering. In particular, beam steering is typically constrained by a maximum steering angle, beyond which it is not possible to generate any beams~\cite{kallnichev2001analysis}. Hence, our analysis can capture the outage probability of ideal beam steering considering the broader sector within which beams can be generated and steered.

\subsection{Coverage Regions}
As the uptilt angle increases, the coverage area changes due to variations in the portions of the UAV corridor occupied by interference areas (I) and beam-served areas without interference (S). To analyze the coverage, we consider 6 cases depending on antenna uptilt angle ($\alpha$) as illustrated in Fig.~\ref{fig:illu}. When we see Fig.~\ref{fig:illu_c2}, green colored area ($\mathsf{S}_1$) and red colored area ($\mathsf{S}_2$) indicate beam-served area from BS-1 and BS-2 without interference, respectively. The green and red overlapped area ($\mathsf{I}_{12}$), orange colored area ($\mathsf{I}_{23}$), and light blue area ($\mathsf{I}_{32}$) represent interference areas between BS-1 and BS-2, between BS-2 and BS-3, and between BS-1 and BS-4, respectively. Note that although BS-3 and BS-4 are not depicted in Fig.~\ref{fig:illu}, their locations correspond to those shown in Fig.~\ref{fig:illu_main}. The red dashed lines from $\mathsf{c}_1$ and from $\mathsf{c}_2$ are coverage borderlines of $\mathsf{I}_{12}$ and $\mathsf{I}_{23}$ regions, which are decided by signal-to-interference-plus-noise-ratio (SINR) threshold for the coverage. Consequently, the coverage area can be described as the combination of beam-served areas without interference (S) and the interference areas (I) contained within the coverage borderlines. These areas are illustrated by the black stripe-patterned regions. Note that while part of the $\mathsf{I}_{12}$ region may experience interference from BS-3 or BS-4, we focus our analysis solely on the interference caused by the most dominant nearby BS.

There are additional height parameters to analyze coverage. Heights from antenna height to the crossing points of beams from BS-1 and BS-2 are $h_3$ (center), and $h_4$ (side), respectively. From \eqref{eq:theta}, we can obtain the relation between $\theta_1$ and $\theta_2$ and between $\theta_1$ and $\theta_3$ as 
\begin{align}\label{eq:theta_1_23}
    \cot(\theta_2)&=\frac{d_1}{h_{\rm x}}-\cot(\theta_1),~\cot(\theta_3)=\frac{d_1}{h_{\rm x}}+\cot(\theta_1).
\end{align}
Then, $h_3$, $h_4$ can be derived as
\begin{align}
    h_3=\frac{d_1}{2}\tan(\alpha),~h_{4}=\frac{d_1}{\cot(\alpha)+\cot(\alpha+\beta)}.
\end{align}

\section{SINR Outage Probability Analysis}\label{sec:cov_ana}
In this section, we define SINR outage probability in the UAV corridor and derive coverage borderlines of interference regions. After that, we derive closed-form expressions of outage probability for unique 6 cases based on the relations between $h_1$, $h_2$, $h_3$, and $h_4$ as shown in Fig.~\ref{fig:illu}.
\subsection{Definition of SINR outage and Coverage Borderlines}
The definition of SINR outage probability can be written as follows:
\begin{align}\label{eq:p_out_def}
    \mathsf{Pr}_{\mathsf{out}}& = 1-\mathsf{Pr}_{\mathsf{in}}=\mathsf{Pr}(\mathsf{SINR}<\tau)~,
\end{align}
where $\tau$ is the SINR threshold and $\mathsf{Pr}_{\mathsf{in}}$ is coverage probability. The distance between a UAV and BS-i can be expressed as
\begin{align}
     R_{i}&=\frac{h_{\rm x}}{\sin(\theta_i)}.
\end{align}
We adopt the free-space path loss model for analysis, which is given as:
\begin{align}
     \mathsf{PL}_{\rm FS}&=\left(\frac{4\pi R_i}{\lambda}\right)^2.
\end{align}
Then, SINR of $\mathsf{I}_{ij}$ interference region can be expressed as 
\begin{align}\label{eq:SINR}
    \mathsf{SINR}_{ij}&=\frac{{\rm K}g_i(\theta_i)/(R_i)^2}{{\rm K}g_j(\theta_j)/(R_j)^2+N_0},\;{\rm K}=\frac{\mathsf{P}_{\mathsf{Tx}}\lambda^2}{16\pi^2},
\end{align}
where $\mathsf{P}_{\mathsf{Tx}}$, $\lambda$, $N_0$ indicate transmit power, wavelength, and noise power, respectively. Based on the geometry in Fig.~\ref{fig:illu_main}, when $i=1$, then $j\in\{2,4\}$, and when $i=2$, then $j\in\{3\}$. These conditions define the interference scenarios in the corridor. Note again that we only consider the left half of the geometry in Fig.~\ref{fig:illu} for our analysis.

The coverage borderline of interference region (I) is the line where SINR becomes threshold $\tau$, which is described in Fig.~\ref{fig:illu} as red dashed lines. First, we can simplify the SINR expression in \eqref{eq:SINR} as
\begin{align}\label{eq:SINR_sp}
    \mathsf{SINR}_{ij}&\approx\frac{{\rm K}G/(R_i)^2}{{\rm K}G/(R_j)^2}=\frac{(R_j)^2}{(R_i)^2},
\end{align}
where the approximation comes from neglecting noise power by considering interference dominant condition. Then, a UAV at the coverage borderline satisfies the following expression,
\begin{align}\label{eq:thr_ij}
    \frac{(R_j)^2}{(R_i)^2}=\tau.
\end{align}
When $h_x=0$, the location of the coverage borderline for $\mathsf{I_{12}}$ satisfies equation from \eqref{eq:thr_ij} as follow:
\begin{align}
    \frac{R_2}{R_1}&=\frac{d_1-d_2}{d_2}=\sqrt{\tau},\nonumber\\
    \to d_2&=\frac{d_1}{\sqrt{\tau}+1},
\end{align}
where $d_2$ indicates the horizontal distance between BS-1 to the coverage borderline for $\mathsf{I_{12}}$ at $h_x=0$ ($\mathsf{c}_1$), which is illustrated in Fig.~\ref{fig:illu_c2}. When $h_x=h_2$, the location of the coverage borderline for $\mathsf{I_{12}}$ holds the following equation:
\begin{align}
    \frac{(R_2)^2}{(R_1)^2}&=\frac{(d_1-d_3)^2+h_2^2}{d_3^2+h_2^2}=\tau,\nonumber\\
    \to d_3&=\frac{d_1-\sqrt{d_1^2-(1-\tau)^2h_2^2-d_1^2(1-\tau)}}{1-\tau},
\end{align}
where $d_3$ indicates the horizontal distance between BS-1 to the coverage borderline for $\mathsf{I_{12}}$ at $h_x=h_2$. Note that this straight coverage borderline by connecting two points at $h_x=0,~h_x=h_2$ gives a good approximation to the exact nonlinear borderline.

In a similar way of $\mathsf{I_{12}}$, when $h_x=0$, the location of the coverage borderline for $\mathsf{I_{23}}$ satisfies the following expression from \eqref{eq:thr_ij}:
\begin{align}
       \frac{R_3}{R_2}&=\frac{d_1+d_4}{d_1-d_4}=\sqrt{\tau},\nonumber\\
       \to d_4&=\frac{(\sqrt{\tau}-1)d_1}{\sqrt{\tau}+1},
\end{align}
where $d_4$ indicates the horizontal distance between BS-1 to the coverage borderline for $\mathsf{I_{23}}$ at $h_x=0$ ($\mathsf{c}_2$ in Fig.~\ref{fig:illu}). When $h_x=h_2$, the location of the coverage borderline for $\mathsf{I_{23}}$ holds following expression:
\begin{align}
       \frac{(R_3)^2}{(R_2)^2}&=\frac{(d_1+d_5)^2+h_2^2}{(d_1-d_5)^2+h_2^2}=\tau\nonumber\\
       \to d_5&=\frac{-(1+\tau)d_1+\sqrt{(1+\tau)^2d_1^2-(1-\tau)^2(h_2^2+d_1^2)}}{1-\tau},
\end{align}
where $d_5$ indicates the horizontal distance between BS-1 to the coverage borderline for $\mathsf{I_{23}}$ at $h_x=h_2$. 

The elevation angles following the linear borderline of $\mathsf{I_{12}}$ at the point $\mathsf{c}_1$ and the linear borderline of $\mathsf{I_{23}}$ at the point $\mathsf{c}_2$ as illustrated in Fig.~\ref{fig:illu_c2} can be expressed by
\begin{align}
    \gamma_1&=\tan^{-1}\left(\frac{h_2}{d_2-d_3}\right),~\gamma_2&=\tan^{-1}\left(\frac{h_2}{d_5-d_4}\right).
\end{align}
The elevation angles at the points $\mathsf{c}_1$ and $\mathsf{c}_2$ to a UAV as described in Fig.~\ref{fig:illu_c2} are given by
\begin{align}\label{eq:theta_c}
    \theta_{\rm c_1}&=\tan^{-1}\left(\frac{h_{\rm x}}{d_2-d_{\rm x}}\right),~\theta_{\rm c_2}=\tan^{-1}\left(\frac{h_{\rm x}}{d_{\rm x}-d_4}\right).
\end{align}
From \eqref{eq:SINR_sp} to \eqref{eq:theta_c}, we derive the distance and angle parameters that help us describe the borderline of the coverage area while analyzing the outage probability.

We summarize the dynamic change of the coverage regions according to the six cases in Fig.~\ref{fig:illu}. In case 1, UAVs in the beam-served area of the BS-1 establish connections with BS-1 in the presence of interference from the BS-2, while a subset of UAVs located within the outage region is served by the BS-2 with interference from the BS-3. In case 2, more UAVs in the outage area are served by the BS-2 as a beam-served area of the BS-2 region without interference from the BS-3 appears. In case 3, the beam-served area of the BS-2 without interference from the BS-3 as well as the beam-served area of the BS-1 without interference from the BS-2 become larger, which increases the number of UAVs within the coverage. In case 4, the beam-served area of the BS-2 without interference from the BS-3 becomes smaller, and the beam-served area of the BS-2 experiencing interference from the BS-3 disappears. In case 5, the beam-served area of the BS-2 disappears and UAVs are only served by BS-1. In case 6, the interference region from the BS-2 disappears.

\subsection{Outage Probability in Case 1 $(h_1 > h_3~\&~h_1 > h_4)$}
Coverage probability in case 1 can be interpreted as the probability that a UAV is located inside the black stripe patterned area in Fig.~\ref{fig:illu_c1}. To calculate this, we divide two integral regions depending on the range of the elevation angle at BS-1 ($\theta_1$): from $h_1$ to the height at $\mathsf{c_3}$ ($h_{\mathsf{c_3}}$) and from $h_{\mathsf{c_3}}$ to $h_{2}$.

We can obtain the relation between $\theta_1$ and $\theta_{\rm c_1}$ and between $\theta_1$ and $\theta_{\rm c_2}$  from \eqref{eq:theta} and \eqref{eq:theta_c} as
\begin{align}\label{eq:theta_1_c}
    \cot(\theta_{\rm c_1})=\frac{d_2}{h_{\rm x}}-\cot(\theta_1),~\cot(\theta_{\rm c_2})=-\frac{d_4}{h_{\rm x}}+\cot(\theta_1).
\end{align}
Then, by plugging in $\theta_1=\alpha + \beta $, $\theta_{\rm c_2}=\gamma_2$, and $h_{\rm x}=h_{\rm c_3}$, we can obtain
\begin{align}
     h_{\rm c_3}&=\frac{d_4}{\cot(\alpha+\beta)-\cot(\gamma_{2})}.
\end{align}
Also, by plugging in $\theta_1=\delta_1$,  $\theta_{\rm c_1}=\gamma_1$ and $\theta_1=\delta_2$,  $\theta_{\rm c_2}=\gamma_2$, we can get
\begin{align}
     \delta_1&=\cot^{-1}\left(\frac{d_2}{h_{\rm x}}-\cot(\gamma_{1})\right),~\delta_2=\cot^{-1}\left(\frac{d_4}{h_{\rm x}}+\cot(\gamma_{2})\right),
\end{align}
where $\delta_1$ and $\delta_2$ denote elevation angles at BS-1 to coverage borderlines of regions $\mathsf{I_{12}}$ and $\mathsf{I_{23}}$, respectively. Then, coverage probability can be expressed as
\begin{align}
    &\mathsf{Pr}_{\rm in}(\tau,\alpha,\beta)=\int_{h_1}^{h_{\rm c_3}}\int_{\delta_1}^{\alpha+\beta} f_{\theta_1}(\theta_1|h_{\rm x})f_{h_{\rm x}}(h_{\rm x})\partial\theta_1\partial h_{\rm x}\nonumber\\
    &+\int_{h_{\rm c_3}}^{h_{2}}\int_{\delta_1}^{\delta_2} f_{\theta_1}(\theta_1|h_{\rm x})f_{h_{\rm x}}(h_{\rm x})\partial\theta_1\partial h_{\rm x},\nonumber\\
    &=\int_{h_1}^{h_{\rm c_3}}\int_{\delta_1}^{\alpha+\beta} \left(\frac{2h_{\rm x}\csc^2(\theta_1)}{d_1}\right)\left(\frac{1}{h_2-h_1}\right)\partial\theta_1\partial h_{\rm x}\nonumber\\
    &+\int_{h_{\rm c_3}}^{h_{2}}\int_{\delta_1}^{\delta_2} \left(\frac{2h_{\rm x}\csc^2(\theta_1)}{d_1}\right)\left(\frac{1}{h_2-h_1}\right)\partial\theta_1\partial h_{\rm x}\nonumber\\
    &=\frac{-h_{\rm c_3}^2+h_1^2}{d_1(h_2-h_1)}\cot(\alpha+\beta)-\frac{(h_1+h_2)}{d_1}\cot(\gamma_1)+\frac{2d_2}{d_1}\nonumber\\
    &+\frac{-h_{2}^2+h_{\rm c_3}^2}{d_1(h_2-h_1)}\cot(\gamma_2)+\frac{-2d_{4}(h_2-h_{\rm c_3})}{d_1(h_2-h_1)}.
\end{align}
The outage probability can then be expressed as
\begin{align}
    \mathsf{Pr}_{\rm out}(\tau,\alpha,\beta)&=1-\mathsf{Pr}_{\rm in}(\tau,\alpha,\beta).
\end{align}
Note that the outage probability is the function of the SINR threshold ($\tau$), the antenna uptilt angle ($\alpha$), and the beamwidth ($\beta$). For simplification, We omit the dependency of $\tau$, $\alpha$, $\beta$ in the outage probability expressions in the remainder of this paper. The outage probability for the other five cases can be derived using a similar approach.

\subsection{Outage Probability in Case 2 $(h_1 > h_3~\&~h_1 < h_4)$}
The coverage area in case 2 in Fig.~\ref{fig:illu_c2} can be divided into four integral regions. By using relations in \eqref{eq:theta_1_23} and \eqref{eq:theta_1_c}, we can obtain height and angle values as follow:
\begin{align}
    \delta_3&=\cot^{-1}\left(\frac{d_1}{h_{\rm x}}-\cot(\alpha)\right),~\delta_4=\cot^{-1}\left(-\frac{d_1}{h_{\rm x}}+\cot(\alpha)\right),\nonumber\\
    h_{\rm c_4}&=\frac{d_1}{\cot(\alpha)},~h_{\rm c_5}=\frac{-d_4-d_1}{\cot(\gamma_{2})-\cot(\alpha)},
\end{align}
where $\delta_3$, $\delta_4$ denote elevation angles at BS-1 to the bottom line of the beam-served area of BS-2 and to the bottom line of the beam-served area of BS-3, respectively, where $h_{\rm c_4}$ and $h_{\rm c_5}$ indicate heights at the points $\mathsf{c_4}$ and $\mathsf{c_5}$, respectively. Then, we derive the outage probability as
\begin{align}
    &\mathsf{Pr}_{\rm out}=1-\int_{h_1}^{h_{4}}\int_{\delta_1}^{\alpha+\beta} f_{\theta_1}(\theta_1|h_{\rm x})f_{h_{\rm x}}(h_{\rm x})\partial\theta_1\partial h_{\rm x}\nonumber\\
    &-\int_{h_{4}}^{h_{\rm c_4}}\int_{\delta_1}^{\delta_3} f_{\theta_1}(\theta_1|h_{\rm x})f_{h_{\rm x}}(h_{\rm x})\partial\theta_1\partial h_{\rm x}\nonumber\\
    &-\int_{h_{\rm c_4}}^{h_{\rm c_5}}\int_{\delta_1}^{\delta_4} f_{\theta_1}(\theta_1|h_{\rm x})f_{h_{\rm x}}(h_{\rm x})\partial\theta_1\partial h_{\rm x}\nonumber\\
    &-\int_{h_{\rm c_5}}^{h_{2}}\int_{\delta_1}^{\delta_2} f_{\theta_1}(\theta_1)f_{h_{\rm x}}(h_{\rm x})\partial\theta_1\partial h_{\rm x}\nonumber\\
    &=1-\frac{-h_{4}^2+h_1^2}{d_1(h_2-h_1)}\cot(\alpha+\beta)+\frac{(h_1+h_2)}{d_1}\cot(\gamma_1)-\frac{2d_2}{d_1}\nonumber\\
    &-\frac{h_{\rm c_4}^2-h_{4}^2}{d_1(h_2-h_1)}\cot(\alpha)+\frac{2(h_{\rm c_4}-h_{4})}{(h_2-h_1)}-\frac{-h_{\rm c_5}^2+h_{\rm c_4}^2}{d_1(h_2-h_1)}\cot(\alpha)\nonumber\\
    &-\frac{2(h_{\rm c_5}-h_{\rm c_4})}{(h_2-h_1)}-\frac{-h_{2}^2+h_{\rm c_5}^2}{d_1(h_2-h_1)}\cot(\gamma_2)+\frac{2d_4(h_{2}-h_{\rm c_5})}{d_1(h_2-h_1)}.
\end{align}

\subsection{Outage Probability in Case 3 $(h_1 < h_3~\&~h_1 < h_4~\&~h_2 > h_{\rm c_4})$}
The coverage area in case 3 in Fig.~\ref{fig:illu_c3} can be divided into six integral regions. By using relations in \eqref{eq:theta_1_c}, we can obtain the height at point $\mathsf{c_6}$ as
\begin{align}
    h_{\rm c_6}&=\frac{d_2-d_1}{\cot(\gamma_{1})-\cot(\alpha)}.
\end{align}
Then, we derive the outage probability as shown at the top of the page.
\begin{figure*}[t]
{\footnotesize
\begin{align}
     &\mathsf{Pr}_{\rm out}=1-\int_{h_1}^{h_{3}}\int_{\alpha}^{\alpha+\beta} f_{\theta_1}(\theta_1|h_{\rm x})f_{h_{\rm x}}(h_{\rm x})\partial\theta_1\partial h_{\rm x}-\int_{h_{3}}^{h_{\rm c_6}}\int_{\delta_3}^{\alpha+\beta} f_{\theta_1}(\theta_1|h_{\rm x})f_{h_{\rm x}}(h_{\rm x})\partial\theta_1\partial h_{\rm x}-\int_{h_{\rm c_6}}^{h_{4}}\int_{\delta_1}^{\alpha+\beta} f_{\theta_1}(\theta_1|h_{\rm x})f_{h_{\rm x}}(h_{\rm x})\nonumber\\
    &\partial\theta_1\partial h_{\rm x}-\int_{h_{4}}^{h_{\rm c_4}}\int_{\delta_1}^{\delta_3} f_{\theta_1}(\theta_1|h_{\rm x})f_{h_{\rm x}}(h_{\rm x})\partial\theta_1\partial h_{\rm x}-\int_{h_{\rm c_4}}^{h_{\rm c_5}}\int_{\delta_1}^{\delta_4} f_{\theta_1}(\theta_1|h_{\rm x})f_{h_{\rm x}}(h_{\rm x})\partial\theta_1\partial h_{\rm x}-\int_{h_{\rm c_5}}^{h_{2}}\int_{\delta_1}^{\delta_2} f_{\theta_1}(\theta_1|h_{\rm x})f_{h_{\rm x}}(h_{\rm x})\partial\theta_1\partial h_{\rm x}\nonumber\\
    &=1-\frac{h_{3}^2-h_1^2}{d_1(h_2-h_1)}(-\cot(\alpha+\beta)+\cot(\alpha))+\frac{h_{\rm c_6}^2-h_3^2}{d_1(h_2-h_1)}(\cot(\alpha+\beta)+\cot(\alpha))-\frac{2(h_{\rm c_6}-h_{3})}{(h_2-h_1)}+\frac{h_{2}^2-h_{\rm c_6}^2}{d_1(h_2-h_1)}(\cot(\gamma_1))\nonumber\\
    &-\frac{2d_2(h_{2}-h_{\rm c_6})}{d_1(h_2-h_1)}+\frac{h_{4}^2-h_{\rm c_6}^2}{d_1(h_2-h_1)}(\cot(\alpha+\beta))-\frac{h_{\rm c_4}^2-h_{4}^2}{d_1(h_2-h_1)}\cot(\alpha)+\frac{2(h_{\rm c_4}-h_{4})}{(h_2-h_1)}+\frac{h_{\rm c_5}^2-h_{\rm c_4}^2}{d_1(h_2-h_1)}(\cot(\alpha))-\frac{2(h_{\rm c_5}-h_{\rm c_4})}{(h_2-h_1)}\nonumber\\
     &+\frac{h_{2}^2-h_{\rm c_5}^2}{d_1(h_2-h_1)}(\cot(\gamma_2))+\frac{2d_4(h_{2}-h_{\rm c_5})}{d_1(h_2-h_1)}
\end{align}}
\hrulefill
\end{figure*}
\subsection{Outage Probability in Case 4 $(h_1 < h_3~\&~h_1 < h_4 < h_2~\&~h_2 < h_{\rm c_4})$}
Following a similar approach as in previous cases, we derve the outage probability in case 4 as follows:
\begin{align}
    &\mathsf{Pr}_{\rm out}=1-\int_{h_1}^{h_{3}}\int_{\alpha}^{\alpha+\beta} f_{\theta_1}(\theta_1|h_{\rm x})f_{h_{\rm x}}(h_{\rm x})\partial\theta_1\partial h_{\rm x}\nonumber\\
    &-\int_{h_{3}}^{h_{\rm c_6}}\int_{\delta_3}^{\alpha+\beta} f_{\theta_1}(\theta_1|h_{\rm x})f_{h_{\rm x}}(h_{\rm x})\partial\theta_1\partial h_{\rm x}\nonumber\\
    &-\int_{h_{\rm c_6}}^{h_{4}}\int_{\delta_1}^{\alpha+\beta} f_{\theta_1}(\theta_1|h_{\rm x})f_{h_{\rm x}}(h_{\rm x})\partial\theta_1\partial h_{\rm x}\nonumber\\
    &-\int_{h_{4}}^{h_{2}}\int_{\delta_1}^{\delta_3} f_{\theta_1}(\theta_1|h_{\rm x})f_{h_{\rm x}}(h_{\rm x})\partial\theta_1\partial h_{\rm x}\nonumber\\
    &=1-\frac{h_{3}^2-h_1^2}{d_1(h_2-h_1)}(-\cot(\alpha+\beta)+\cot(\alpha))\nonumber\\
    &+\frac{h_{\rm c_6}^2-h_3^2}{d_1(h_2-h_1)}(\cot(\alpha+\beta)+\cot(\alpha))-\frac{2(h_{\rm c_6}-h_{3})}{(h_2-h_1)}\nonumber\\
    &+\frac{h_{2}^2-h_{\rm c_6}^2}{d_1(h_2-h_1)}(\cot(\gamma_1))-\frac{2d_2(h_{2}-h_{\rm c_6})}{d_1(h_2-h_1)}\nonumber\\
    &+\frac{h_{4}^2-h_{\rm c_6}^2}{d_1(h_2-h_1)}(\cot(\alpha+\beta))-\frac{h_{2}^2-h_{4}^2}{d_1(h_2-h_1)}\cot(\alpha)\nonumber\\
    &+\frac{2(h_{2}-h_{4})}{(h_2-h_1)}.
\end{align}
\subsection{Outage Probability in Case 5 $(h_1 < h_3 < h_2 ~\&~h_2 < h_4$)}
Outage probability in case 5 can be expressed as
\begin{align}
    &\mathsf{Pr}_{\rm out}=1-\int_{h_1}^{h_{3}}\int_{\alpha}^{\alpha+\beta} f_{\theta_1}(\theta_1|h_{\rm x})f_{h_{\rm x}}(h_{\rm x})\partial\theta_1\partial h_{\rm x}\nonumber\\
    &-\int_{h_{3}}^{h_{\rm c_6}}\int_{\delta_3}^{\alpha+\beta} f_{\theta_1}(\theta_1|h_{\rm x})f_{h_{\rm x}}(h_{\rm x})\partial\theta_1\partial h_{\rm x}\nonumber\\
    &-\int_{h_{\rm c_6}}^{h_{2}}\int_{\delta_1}^{\alpha+\beta} f_{\theta_1}(\theta_1|h_{\rm x})f_{h_{\rm x}}(h_{\rm x})\partial\theta_1\partial h_{\rm x}\nonumber\\
    &=1-\frac{h_{3}^2-h_1^2}{d_1(h_2-h_1)}(-\cot(\alpha+\beta)+\cot(\alpha))\nonumber\\
    &+\frac{h_{\rm c_6}^2-h_3^2}{d_1(h_2-h_1)}(\cot(\alpha+\beta)+\cot(\alpha))-\frac{2(h_{\rm c_6}-h_{3})}{(h_2-h_1)}\nonumber\\
     &+\frac{h_{2}^2-h_{\rm c_6}^2}{d_1(h_2-h_1)}(\cot(\alpha+\beta)+\cot(\gamma_1))-\frac{2d_2(h_{2}-h_{\rm c_6})}{d_1(h_2-h_1)}.
\end{align}
\subsection{Outage Probability in Case 6 $(h_2 < h_3 ~\&~h_2 < h_4$)}
Outage probability in case 6 can be derived as
\begin{align}
    &\mathsf{Pr}_{\rm out}=1-\int_{h_1}^{h_{2}}\int_{\alpha}^{\alpha+\beta} f_{\theta_1}(\theta_1|h_{\rm x})f_{h_{\rm x}}(h_{\rm x})\partial\theta_1\partial h_{\rm x}\nonumber\\
    &=1-\frac{h_{2}+h_1}{d_1}(-\cot(\alpha+\beta)+\cot(\alpha)).
\end{align}

\section{Cosine Beam Shape and Air-to-Ground Path Loss Model for Monte Carlo Simulations}
In previous sections, we adopted simplified assumptions, such as using a simple rectangular-shaped beam pattern in \eqref{eq:bp_model} and a free-space path loss model, to facilitate tractable analysis. However, these assumptions may not adequately capture the realistic propagation characteristics of the UAV corridor environment. To address this limitation, we introduce a more realistic cosine beam pattern, which better represents the main lobe shape of practical antenna patterns. Additionally, we incorporate an air-to-ground path loss model that takes into account the line-of-sight (LoS) and non-line-of-sight (NLoS) probabilities based on the elevation angle of a UAV-BS link, considering the presence of obstacles that can obstruct the LoS path. By employing the cosine beam pattern and the air-to-ground path loss model in our simulations, we aim to observe whether the trends in coverage probability align with the analysis conducted. This allows us to validate the consistency between the analytical results from the simplistic system model in Section~\ref{sec:system} and the outcomes of the more realistic system model, which is obtained by Monte Carlo simulations.

The cosine beam pattern model is given by~\cite{maeng2023base,yu2017coverage}:
\begin{align}
    g_{\rm i}(\theta_{\rm i})&=
    \begin{cases}\label{eq:bp_model_cos}
    N_{\rm t}\cos^2\left(\frac{\pi N_{\rm t}}{2}x\right) & |x|\leq \frac{1}{N_{\rm t}}\\
    0 & \text{otherwise}
    \end{cases}~,
\end{align}
where $x=\left(\cos(\theta_{\rm i})-\cos(\alpha+\beta/2)\right)/2$ when we consider half-wavelength antenna spacing, and $N_{\rm t}$ denotes the number of antenna elements. The air-to-ground path loss model with probabilistic LoS is expressed as~\cite{al2014optimal,al2017modeling}  
\begin{align}\label{eq:pl_avg}
\mathsf{PL}_{\rm A2G}&=\textbf{P}_{\rm LoS}(\eta_{\rm LoS}\mathsf{PL}_{\rm FS})+\textbf{P}_{\rm NLoS}(\eta_{\rm NLoS}\mathsf{PL}_{\rm FS})~,
\end{align}
where $\eta_{\rm LoS}$ and $\eta_{\rm NLoS}$  denote the excessive path loss of the LoS and the NLoS links, respectively, while $\textbf{P}_{\rm LoS}$, $\textbf{P}_{\rm NLoS}$ indicate LoS and NLoS probabilities, respectively. 

\section{Numerical Results} \label{sec:results}

\begin{table}[t]
\renewcommand{\arraystretch}{1.1}
\caption{Simulation settings for UAV corridor coverage analysis.}
\label{table:settings}
\centering
\begin{tabular}{lc}
\hline
Parameter & Value \\
\hline\hline
Transmit power ($\mathsf{P}_\mathsf{Tx}$) & $30$~dBm\\
Horizontal distance between BSs ($d_1$) & $1000$~m \\
Minimum UAV corridor height from a BS height ($h_1$) & $100$~m \\
Maximum UAV corridor height from a BS height ($h_2$) & $300$~m \\
SINR threshold for coverage ($\tau$) & $2$~dB \\
Maximum antenna gain ($G$) & $297.6$/$\beta$~dB~\cite{venugopal2016device} \\
Carrier frequency & $3$~GHz \\
Bandwidth (BW) & $20$~MHz\\
Thermal noise (TN) & $-174$~dBm/Hz\\
Noise figure (NF) & $9$~dB\\\hline
\hline
\end{tabular}
\vspace{-0.2in}
\end{table}

\begin{figure}[t!]
    \centering
    \subfloat[$d_1=1000$~m.]{\includegraphics[width=0.48\textwidth]{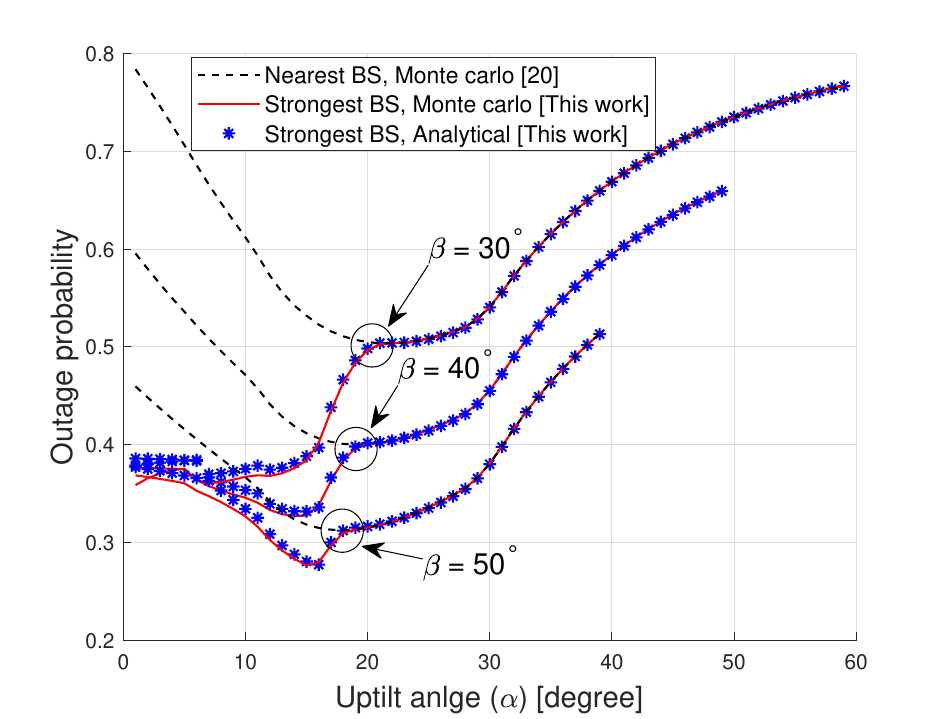}\label{fig:p_out_a_1}}~
    \subfloat[$d_1=1300$~m]{\includegraphics[width=0.48\textwidth]{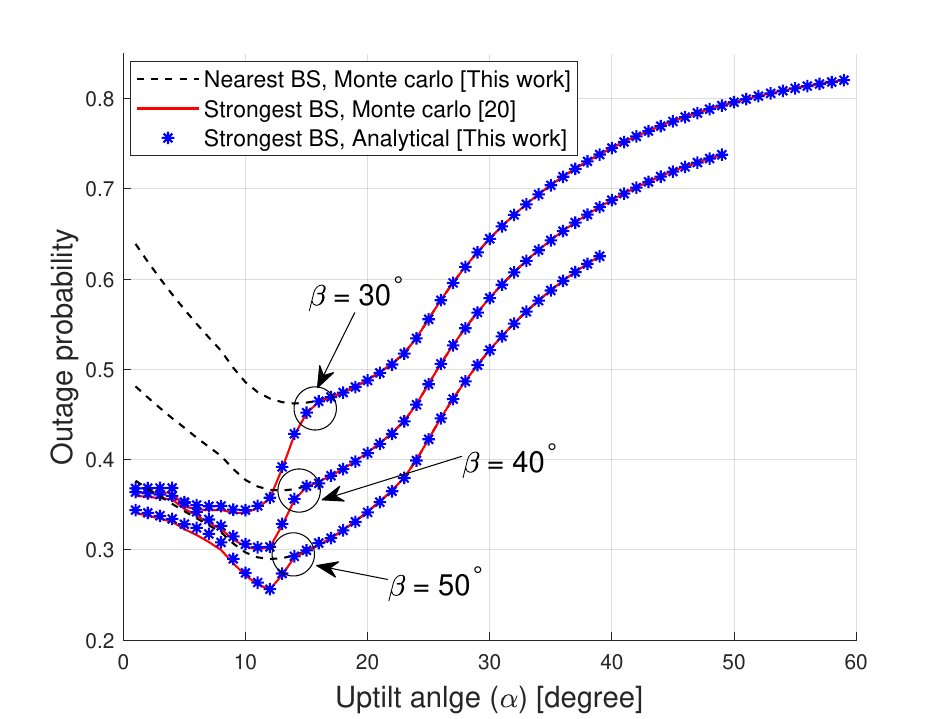}\label{fig:p_out_a_2}}
    \caption{Outage probability depending on uptilt angle $\alpha$ based on analysis model. The strongest BS association outperforms the nearest BS association analyzed in \cite{maeng2023base} with respect to the minimum outage probability. Outage probabilities are convex curves with respect to uptilt angles.}\label{fig:p_out_a}
\end{figure}

\begin{figure}[t!]
    \centering
    \subfloat[$d_1=1000$~m.]{\includegraphics[width=0.48\textwidth]{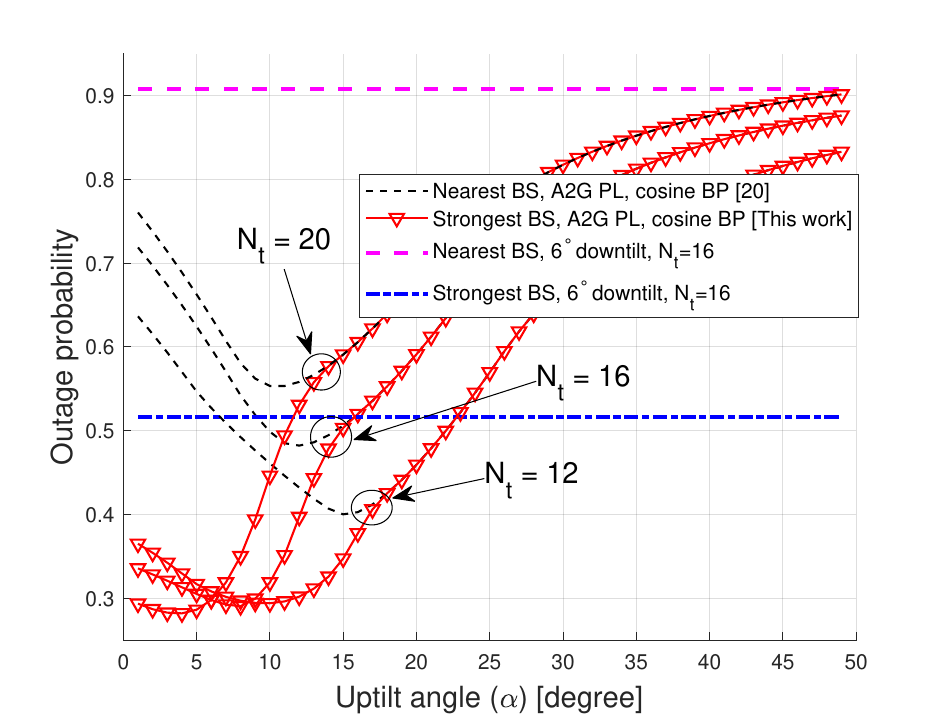}\label{fig:p_out_b_1}}~
    \subfloat[$d_1=1300$~m.]{\includegraphics[width=0.48\textwidth]{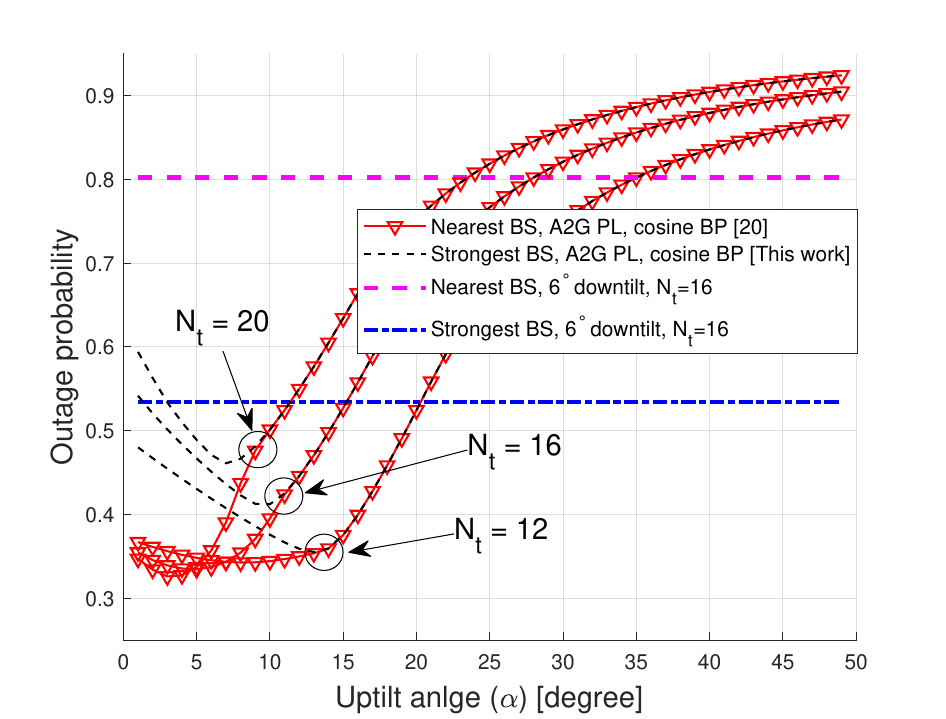}\label{fig:p_out_b_2}}
    \caption{Outage probability depending on uptilt angle $\alpha$ based on realistic models with air-to-ground path loss and cosine beam
pattern. The optimum uptilt angles are lower with more realistic models when the strongest association is considered.}\label{fig:p_out_b}
\end{figure}

\begin{figure}[t!]
    \centering
    \subfloat[$\alpha=8^{\circ}$ (case 2)]{\includegraphics[width=0.45\textwidth]{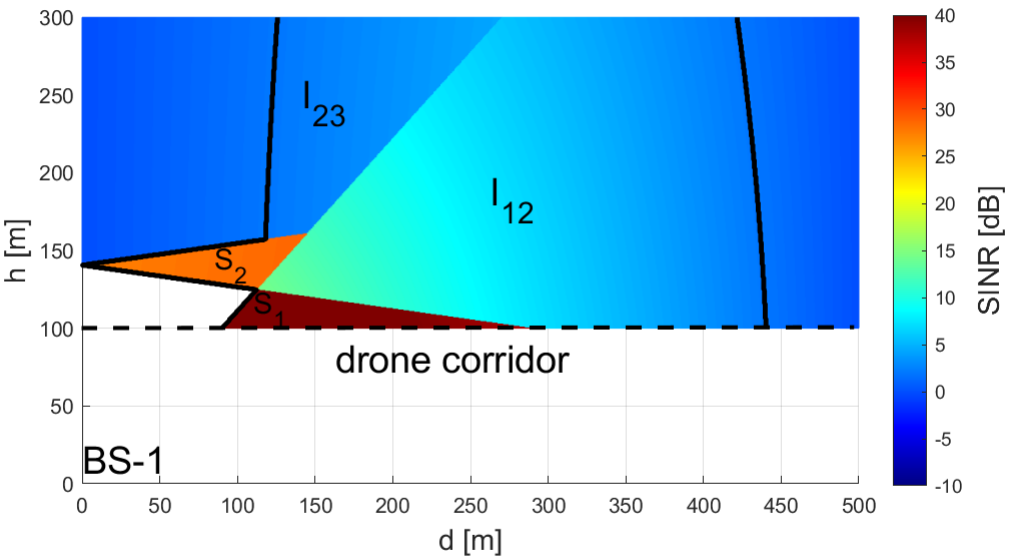}}\vspace{-0.08in}

    \subfloat[$\alpha=13^{\circ}$ (case 3)]{\includegraphics[width=0.45\textwidth]{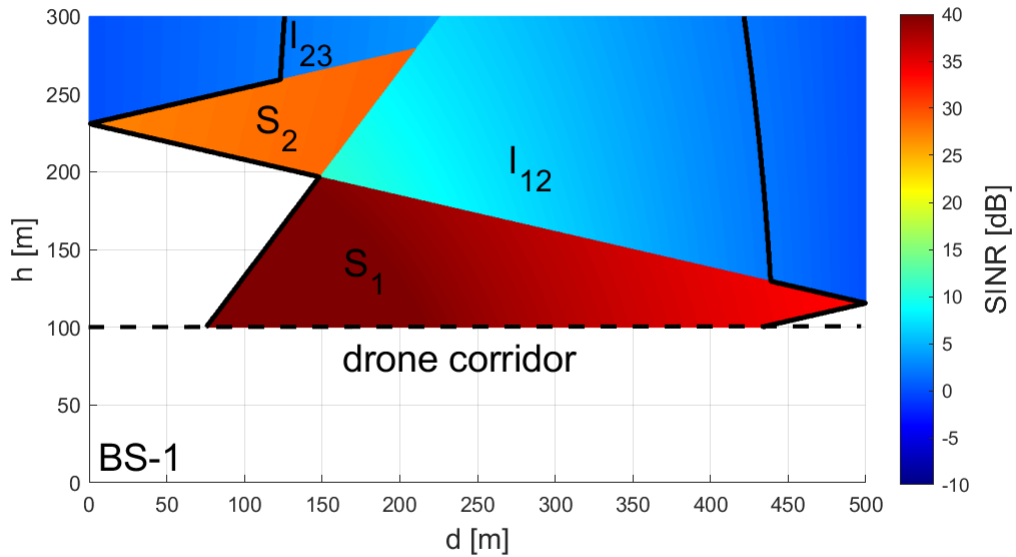}}\vspace{-0.08in}

    \subfloat[$\alpha=17^{\circ}$ (case 4)]{\includegraphics[width=0.45\textwidth]{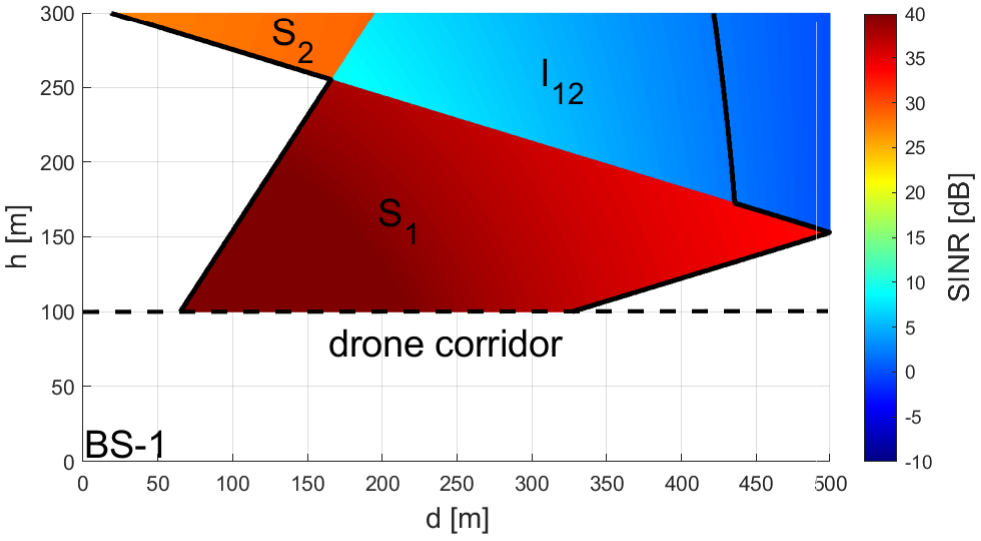}}\vspace{-0.08in}

    \subfloat[$\alpha=25^{\circ}$ (case 5)]{\includegraphics[width=0.45\textwidth]{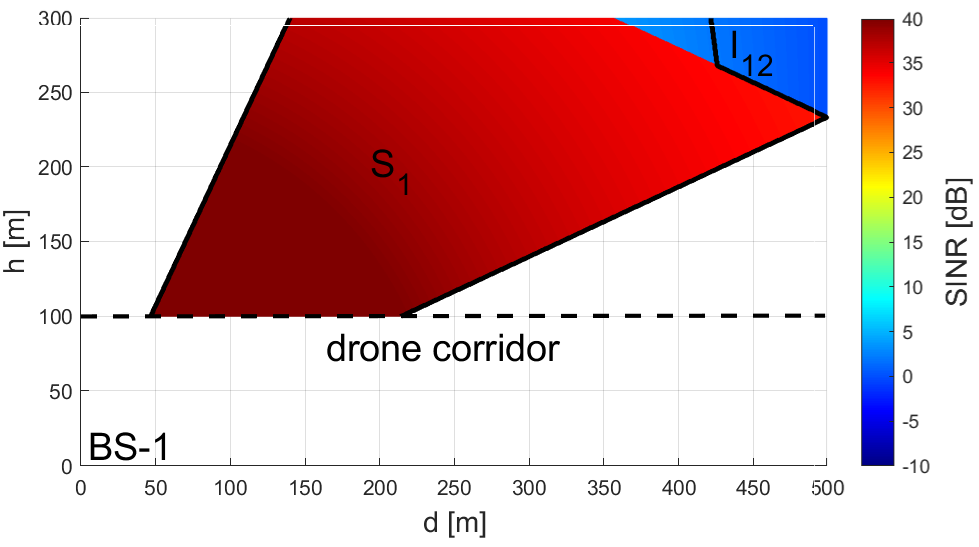}}
    \caption{SINR heatmap of BS-1 serving area depending on uptilt angle $\alpha$. The coverage area is inside the black solid line.  Representative uptilt angles show Case 2 - 5 in Fig.~\ref{fig:illu}. SINR threshold $\tau=2$ dB, beamwidth $\beta=40^{\circ}$, the minimum UAV corridor height ($h_1$) = 100~m, and the maximum UAV corridor height ($h_2$) = 300~m.}\label{fig:SINR_heatmap}
\end{figure}

In this section, we present simulation results to verify the analysis of the coverage probability and compare the performance of the strongest BS association with the nearest BS association, which is investigated in~\cite{maeng2023base}. The parameter setups are listed in Table~I. Noise power in \eqref{eq:SINR} is calculated by $N_0=(\text{TN})+10\log_{10}(\text{BW})+(\text{NF})~\text{[dBm]}$.

Fig.~\ref{fig:p_out_a} show the outage probability as a function of the antenna uptilt angle, the beamwidth, and the distance between BSs. We also compare the strongest BS association of this paper with the nearest BS association in \cite{maeng2023base}. It is observed that the strongest BS association outperforms the nearest BS association. This is because of the extended coverage achieved through the connection with BS-2 in the strongest BS association. This observation highlights the advantage of considering the strongest BS association for improved coverage for a given UAV corridor geometry. We also observe that both associations result in convex curves in terms of the antenna uptilt angle. This suggests that there exists an optimal uptilt angle that maximizes coverage. Finding this optimal angle is crucial for achieving the best possible coverage performance in UAV corridors. In addition, it is observed that as the distance between BSs ($d_1$) increases, the minimum outage probability becomes lower. As the beamwidth ($\beta$) decreases, the performance gap between the strongest BS association and the nearest BS association becomes wider.

In Fig.~\ref{fig:p_out_b}, we employ a realistic propagation model incorporating an air-to-ground path loss model with a probabilistic LoS link, and a cosine-shaped beam pattern. The simulation results demonstrate that the strongest BS association continues to outperform the nearest BS association even when evaluated using these realistic models. It is observed that the strongest BS association archives better performance than the nearest BS association by the realistic models. Additionally, we observe that the minimum outage probability achieved by the strongest BS association remains similar across different antenna setups, indicating robust performance across varying system configurations. Furthermore, as the distance between BSs ($d_1$) increases the minimum outage probability becomes slightly higher, which is in contrast to the results with the simpler mathematical models in Fig.~\ref{fig:p_out_a}. On the other hand, the performance gap between the strongest BS association and the nearest BS association becomes wide similar to the behavior in Fig.~\ref{fig:p_out_a}. Moreover, we include simulation results with the downtilted antenna scenario to compare it with the uptilted antenna scenario. We consider the same realistic propagation model with a 6-degree downtilt angle, which is a typical angle to serve ground users in cellular networks~\cite{maeng2023base,chowdhury2021ensuring}. We observe that in both the nearest BS and the strongest BS associations, the performance of the downtilted antenna scenario is worse than the uptilted antenna scenario, which implies that the serving area of the UAV corridor from the downtilted main beam and the ground reflection signal is smaller than the case that we consider with the uptilted antenna.

Fig.~\ref{fig:SINR_heatmap} shows the SINR heatmap for the serving area of BS-1 in a UAV corridor, considering the left half of corridor area, with the antenna uptilt angle $\alpha$ as the varying parameter. The black solid lines represent the SINR values that meet the coverage threshold $\tau$ in dB, which delineates the boundaries of the coverage area. We can clearly see the beam-served areas without interference $\mathsf{S_1}$, $\mathsf{S_2}$, interference area $\mathsf{I_{12}}$, $\mathsf{I_{23}}$, and coverage borderlines. As the uptilt angle increases, we can observe coverage area changes by case 2 - 5 in Fig.~\ref{fig:illu}. The SINR is abruptly changed as the beam-served region without interference ($\mathsf{S_1}$) is translated into the interference region ($\mathsf{I_{12}}$). This is due to the fact that interference from the BS-2 highly decreases SINR when the noise power level is significantly lower when compared with the interference power level.

\section{Discussion}\label{sec:discussion}
Aside from the terrestrial BS networks, a high-altitude platform station (HAPS) can be considered to serve a UAV corridor. HAPS is a non-terrestrial and aerostatic network node that operates in the stratosphere and provides communication services to the ground~\cite{kurt2021vision,arum2020review}. If HAPS is utilized in serving the UAV corridor, a single HAPS can cover a wider range of areas of the UAV corridor compared with a terrestrial BS. In addition, with the flexible deployment of the HAPS nodes, the outage area can be efficiently managed. For instance, the coverage can be optimized by adjusting the altitude of HAPS nodes depending on the beamwidth while the coverage is optimized by the antenna uptilt angle. However, a HAPS is required to consider the maximum transmit power constraint, high computational capacity, more demanding operational constraints, higher operational costs, and short mission durations that depend on the power harvested at the solar panels. Moreover, the backhaul constraint and control signal from the terrestrial networks can limit the performance of a HAPS. Furthermore, a HAPS should coordinate with the terrestrial networks to manage the interference to the ground users while the terrestrial BS networks with antenna uptilt merely interfere with the ground users.

As expected, our results show that the uptilted antennas at the BSs provide better coverage to the UAV corridor than the downtilted antennas that are optimized for serving to the ground users. In our prior work~\cite{maeng2023base}, we consider the downtilted antenna scenario to serve the UAV corridor with the \emph{nearest BS association rule} and simulation results show that the uptilted antennas achieve a higher average SINR compared with the downtilted antennas. In addition, in \cite{chowdhury2021ensuring}, we consider typical hexagonal multicell networks to serve UAVs, and ground base stations are equipped with both uptilted and downtilted antennas. In simulation results, we show that the BSs with the uptilted antennas achieve a higher signal-to-interference ratio (SIR) than BSs only with the downtilted antennas. Moreover, the interference from ground reflection by the downtilted antennas reduces the SIR when the BS is equipped with both uptilt and downtilted antennas.

\section{Concluding Remark}\label{sec:conclusion}
In this paper, we have investigated the impact of antenna uptilt angle on coverage improvement in UAV corridor networks. Our analysis focuses on understanding the changes in beam-served areas and interference areas as the uptilt angle varies, leading to the division of the analysis into six different cases. To quantify the coverage performance, we have derived closed-form expressions for the outage probability in each of the identified cases. This allows us to mathematically characterize the probability of a UAV experiencing an outage in the network that is designed to serve the UAV corridor with a minimum SINR guarantee. The UAV may also be served by the same network with a lower SINR at outage regions, or by other networks with no SINR guarantee.  The simulation and analytical results demonstrate that the strongest BS association strategy achieves superior coverage compared to the nearest BS association when the antenna uptilt angle is set to its optimal value. We also observe that the outage probability exhibits a convex curve in relation to the uptilt angle, suggesting the existence of an optimal angle that maximizes coverage performance.

\bibliographystyle{IEEEtran} 
\bibliography{IEEEabrv,bibfile}

\end{document}